\documentclass[12pt,epsf]{article}
\usepackage{graphicx}

\addtocounter{page}{0} \tolerance=5000

\newcommand{\beq}{\begin{equation}}
\newcommand{\eeq}{\end{equation}}
\newcommand{\beqn}{\begin{eqnarray}}
\newcommand{\eeqn}{\end{eqnarray}}

\newcommand{\vA}{\mathbf{A}}

\newcommand{\E}{\mathbf{E}}

\newcommand{\F}{\mathbf{F}}
\newcommand{\bH}{\mathbf{H}}

\newcommand{\bj}{\mathbf{j}}
\newcommand{\bk}{\mathbf{k}}

\newcommand{\n}{\mathbf{n}}

\newcommand{\bq}{\mathbf{q}}
\newcommand{\br}{\mathbf{r}}

\newcommand{\s}{\mathbf{s}}

\newcommand{\bv}{\mathbf{v}}

\newcommand{\ga}{\mbox{${\gamma}$}}

\newcommand{\de}{\mbox{${\delta}$}}
\newcommand{\De}{\mbox{${\Delta}$}}

\newcommand{\ep}{\mbox{${\varepsilon}$}}
\newcommand{\om}{\mbox{${\omega}$}}

\newcommand{\Na}{\mbox{{\boldmath$\nabla$}}}
\newcommand{\si}{\mbox{{\boldmath$\sigma$}}}

\newcommand{\na}{\mbox{${\nabla}$}}
\newcommand{\pa}{\mbox{${\partial}$}}

\begin{document}
\begin{center}
{\Large \bf Cerenkov radiation of spinning particle}
\end{center}

\vspace{0.7cm}

\begin{center}
I.B. Khriplovich\footnote{khriplovich@inp.nsk.su}
\\Budker
Institute of Nuclear Physics\\ 630090 Novosibirsk, Russia,\\ and
Novosibirsk University
\end{center}

\vspace{0.7cm}

\begin{abstract}
The Cerenkov radiation of a neutral particle with magnetic moment
is considered, as well as the spin-dependent contribution to the
Cerenkov radiation of a charged spinning particle. The
corresponding radiation intensity is obtained for an arbitrary
value of spin and for an arbitrary spin orientation with respect
to velocity.
\end{abstract}


\vspace{0.7cm}

{\bf 1.} The problem of Cerenkov radiation of a neutral particle with magnetic moment,
moving in a medium with the refraction index $n$ with velocity $v
> c/n$, was considered previously in Refs. [1-6]. The
magnetic dipole was modeled therein classically, either by a loop
with current, or by a pair of magnetic monopole -- antimonopole.
Thus obtained results are rather model-dependent, and the
conclusion made in Ref. \cite{fr2} is that the situation with the
problem of Cerenkov radiation by a magnetic moment is not exactly
clear.

In the present work the problem is addressed as follows. A
spinning particle, charged or neutral, with magnetic moment is
treated as a point-like one, i.e. it is described by a
well-localized wave packet. As to the spin $s$, it has an
arbitrary half-integer or integer value, starting with $s = 1/2$.
In particular, in the limit $s \gg 1$ we arrive at the classical
internal angular momentum and classical magnetic moment. The
result obtained below for a neutral particle with magnetic moment
differs considerably from all previous ones. As to the
spin-dependent contribution to the Cerenkov radiation of a charged
particle, I am not aware of any previous results for it.

Certainly, the effects analyzed here are tiny, too small perhaps
to be observed experimentally. Hopefully however, their
investigation is of some theoretical interest.

\vspace{5mm}

{\bf 2.} We start with the electric and magnetic fields created by
a point-like neutral particle with magnetic moment $e\,\s\,g/(2m)
= (es\, g/(2m))\,\si$; here and below $g$ is the $g$-factor, and
$\si = \s/s$. Of course, for $s = 1/2$, vector $\si$ consists of
the common spin $\sigma$-matrices, and in the classical limit $s
\gg 1$, $\si$ is just a unit vector directed along $\s$. In the
particle rest frame, the four-dimensional current density is
\beq\label{rf}
j^{\,(rf)}_{\alpha} = (0, \bj^{\,(rf)}) = \frac{e s\, g}{2m}
\left(0, \Na \times \si^{\,(rf)} \right)\, \de(\br^{\,(rf)})\,.
\eeq
In the laboratory frame, we are working in, this
Lorentz-transformed current looks formally as follows:
\[
j_{\alpha} = \left(\gamma v(\n \bj^{\,(rf)}), \; \bj^{\,(rf)} -
\n(\n\bj^{\,(rf)}) + \gamma\, \n(\n\bj^{\,(rf)})\right);
\quad \gamma = 1/\sqrt{1-v^2}, \quad \n = \bv/v
\]
(we put throughout $c=1$). Now, we have to go over in
$\bj^{\,(rf)}$ from the rest-frame coordinates $\br^{\,(rf)}$ to the laboratory ones:
\[
\br^{\,(rf)} = \left(\gamma (x - v t),\, y,\, z \right).
\]
Under this Lorentz transformation,
\[
\de(\br^{\,(rf)}) = \de(\gamma (x - v t))\,\de(y)\,\de(z) =
\frac{1}{\gamma}\,\de(x - v t)\,\de(y)\,\de(z)=
\frac{1}{\gamma}\,\de(\br - \bv t)\,.
\]
Besides this overall factor $1/\gamma$, the components of gradient
transform obviously as follows:
\[
\na^{\,(rf)}_{x}\,\de(\br - \bv t) =
\frac{1}{\gamma}\,\na_{x}\,\de(\br - \bv t),\quad
\na^{\,(rf)}_{y,\,z}\,\de(\br - \bv t) = \na_{y,\,z}\,\,\de(\br -
\bv t).
\]
As to the spin operators $\si^{\,(rf)}$, also entering $\bj^{\,(rf)}$, their transformation
law is the same as that for $\bj^{\,(rf)}$ itself:
\[
\si = (\sigma_x\,,\,\sigma_y\,,\,\sigma_z)= \si^{\,(rf)} -
\n(\n\si^{\,(rf)}) + \gamma\, \n(\n\si^{\,(rf)}) =
(\gamma\,\sigma^{\,(rf)}_{x}\,,\,\sigma^{\,(rf)}_{y}\,,\,\sigma^{\,(rf)}_{z})\,,
\]
or
\[
\sigma^{\,(rf)}_{x} = \frac{1}{\gamma}\,\sigma_{x},\quad
\sigma^{\,(rf)}_{y,\,z} = \sigma_{y,\,z}\,.
\]

Thus, in the laboratory frame the four-dimensional current density,
created by the magnetic moment $(es g/(2m))\,\si$, is \footnote{Here and below
$(\si \bv \Na) = \si \cdot [\bv \times \Na] = [\si \times \bv]\cdot \Na$, etc.}
\beq\label{lf}
j_{\alpha}^{\,g}(\br, t) = \frac{es\, g}{2m} \left((\si \bv \Na),
\; (1-v^2)\Na \times \si + \bv (\si \bv \Na)\right)\, \de(\br -
\bv t).
\eeq
{\bf We note that this 4-current density, as well as the initial rest-frame one 
(\ref{rf}), is orthogonal to the 4-velocity $u_{\alpha}$:
$u_{\alpha}j_{\alpha}=0\,.$ This is an extra check of the above transformations.
Let us note also that the current density (\ref{lf})} can be conveniently rewritten as the sum of
two four-currents, each of them being conserved by itself:
\beq\label{lf1}
j_{\alpha}^{\,1g}(\br, t) = \frac{es\, g}{2m}\, (\si \bv \Na)
\left ( 1,\; \bv \right)\, \de(\br - \bv t)\,,
\eeq
\beq\label{lf2}
j_{\alpha}^{\,2g}(\br, t) = \frac{es\, g}{2m}\,(1-v^2) \left(0, \;
\Na \times \si \right)\, \de(\br - \bv t)\,.
\eeq

We are interested in the back-reaction of the field created by the
current (\ref{lf}) upon the spin of the particle. This interaction
is
\beq\label{ug}
H_g = \int d\br j_{\alpha}^{\,g}(\br - \bv t) A_{\alpha}(\br) =
\frac{es g}{2m}\,\si\left[\bH -
\frac{\gamma}{\gamma+1}\;\bv(\bv\bH) - \bv \times \E \right],
\eeq
where both field strengths, $\bH$ and $\E$, are taken at the point
of spin location $\br = \bv t$. This is the usual interaction of
the magnetic moment of a relativistic neutral particle with an
external electromagnetic field. In fact, we have omitted in the
final expression a term proportional to the total derivative of
the vector potential, $d\vA/dt = \pa\vA/\pa t + (\bv \Na) \vA$,
since a total time derivative in interaction does not result at
all in observable effects. Moreover, in the present case the
vector potential $\vA$, together with the current creating it,
depends on the combination $\br - \bv t$ only, so that this total
derivative vanishes identically.

This line of reasoning is generalized easily for the case of a
charged particle. To this end one has to supplement the spin
current (\ref{lf}) with the following, also conserved, contribution:
\beq\label{tc}
j_{\alpha}^{\,th}(\br, t) =
-\,\frac{es}{m}\,\frac{\gamma}{\gamma+1}\,(\si \bv \Na) \left(1,
\; \bv \right)\, \de(\br - \bv t).
\eeq
In its turn, this current generates one more contribution to the
spin interaction with electromagnetic field:
\beq\label{ut}
H_{th} = \int d\br j_{\alpha}^{\,th}(\br - \bv t)
A_{\alpha}^{}(\br) = \frac{es}{m}\,\si\left[\left(1
-\frac{1}{\gamma}\right)\bH - \frac{\gamma}{\gamma+1}\;\bv(\bv\bH)
- \,\frac{\gamma}{\gamma+1}\,\bv \times \E \right],
\eeq
which describes the well-known Thomas precession. In this
expression we have omitted as well, and by the same reasons, a
term proportional to the total derivative $d\vA/dt = \pa\vA/\pa t
+ (\bv \Na) \vA$. Finally, from now on, we will work with the
following total interaction
\beq\label{uto}
H = H_g + H_{th} = - \frac{es}{2m}\;\si\left[\left(g - 2
+\frac{2}{\gamma}\right)\bH - (g -
2)\frac{\gamma}{\gamma+1}\;\bv(\bv\bH) - \left(g -
\frac{2\gamma}{\gamma+1}\right) \bv \times \E \right],
\eeq
and the total spin current
\[
j_{\alpha}(\br, t) = j_{\alpha}^{\,g}(\br, t) +
j_{\alpha}^{\,th}(\br, t)
\]
\beq\label{sto}
= \frac{es}{2m} \left\{\left(g -
\frac{2\gamma}{\gamma+1}\right)(\si \bv \Na) \left(1, \; \bv
\right)\; + g (1-v^2)\left(0, \; \Na \times \si\right)\right\}\,
\de(\br - \bv t).
\eeq

Hamiltonian (\ref{uto}) not only generates the spin precession, including
of course the Thomas effect. It produces as well the relativistic Stern-Gerlach force
\beq\label{f}
\F = - \Na H.
\eeq
Obviously, this force results in the energy loss and therefore is
antiparallel to the velocity $\bv$ of the spinning particle. Thus,
the energy loss per unit time, or the (positive) radiation
intensity, is
\beq\label{I0}
I = - \F\bv = (\bv\Na)H.
\eeq

Let us note here that the field strengths $\bH$, $\E$, being
created by the current density $j_{\alpha}(\br, t)$, depend
themselves on the non-commuting operators $\si$. Therefore, to
guarantee that expression (\ref{I0}) is hermitian, one should,
strictly speaking, properly symmetrize the products of
$\sigma$-operators therein. In fact, however, the final result
(see (\ref{gr}) below) proves to be hermitian
automatically, without extra efforts.

\vspace{5mm}

{\bf 3.} The derivation in this section, resulting in general
expression (\ref{gr}) (see below) for the spectral intensity, follows
essentially that applied in Ref.~\cite{ll} to the problem of the
common Cerenkov radiation.

We will calculate the radiation intensity by going over to the
Fourier  transforms $\bH_\bk$ and $\E_\bk$ of the field strengths,
defined as follows:
\[
\bH(\br - \bv t) = \int d^3 k\,e^{i\bk(\br - \bv t)}\,\bH_\bk,
\quad \E(\br - \bv t) = \int d^3 k\, e^{i\bk(\br - \bv
t)}\,\E_\bk.
\]
For our purpose, the wave vectors $\bk$ are conveniently decomposed
into the components parallel to the velocity $\bv$ and orthogonal
to it: $\bk = \bq + \n\,\omega/v$, $\om = \bk\bv$, $(\bq\bv) =0$.
Then, at the position of the point-like source we have
\beq\label{NH}
(\bv\Na)\,\bH(\br = \bv t) = \int d^3 k\,i\om \,\bH_\bk = -
\frac{1}{v}\int d^2 q \int^{\infty}_{-\infty}d\om \,\om \,\bk
\times \vA_{\bk},
\eeq
\beq\label{EH}
(\bv\Na)\,\E(\br = \bv t) = \int d^3 k\,i\om \,\E_\bk = -
\frac{1}{v}\int d^2 q \int^{\infty}_{-\infty}d\om \,\om \,(\om
\vA_{\bk} - \bk \phi_{\bk}),
\eeq
where $\phi_{\bk}$ and $\vA_{\bk}$ are the Fourier transforms of
the electromagnetic scalar and vector potentials.

In the generalized Lorenz gauge
\[
{\rm div} \vA + \frac{\pa \hat{\ep} \phi}{\pa t} =0\,,
\]
the wave equations for potentials are
\beq\label{p}
\hat{\ep}\left(\De \phi\, -\, \hat{\ep}\,\frac{\pa^2 \phi}{\pa
t^2}\right) = \,-\, \,4\pi j_0\,(\br - \bv t)\,=\,-\, \,4\pi
\frac{es}{2m}\left(g - \frac{2\gamma}{\gamma+1}\right)(\si \bv \Na)\,\de(\br - \bv t)\,,
\eeq
\[
\De \vA\, -\, \hat{\ep}\,\frac{\pa^2 \vA}{\pa t^2} = \,-\,4\pi
\bj\,(\br - \bv t)\,
\]
\beq\label{A}
=\,-\,4\pi \frac{es}{2m} \left\{\left(g -
\frac{2\gamma}{\gamma+1}\right)(\si \bv \Na) \bv \; + g (1-v^2)\;
\Na \times \si\right\}\, \de(\br - \bv t).
\eeq
Here the ``dielectric constant'' $\hat{\ep}$ should be understood
as an operator; we use below its Fourier-transform $\ep(\om)$. As
to the permeability $\mu(\om)$, for the frequencies of interest to
us, it can be put equal to unity.

Now, for the Fourier transforms of the potentials we obtain
\[
\phi_{\bk}\,=\, \frac{i}{2 \pi^2}\,\frac{1}{\ep(\om)}
\frac{es}{2m}\left(g -
\frac{2\gamma}{\gamma+1}\right)\frac{(\bv\bk\si)}{k^2 - \ep(\om)\,
\om^2}\quad\quad\quad\quad\quad\quad
\]
\beq\label{pt}
=\, \frac{i}{2 \pi^2}\,\frac{1}{\ep(\om)} \frac{es}{2m}\left(g -
\frac{2\gamma}{\gamma+1}\right)\;\frac{(\bv\bq\si)}{q^2 -
[\ep(\om)\,- 1/v^2]\,\om^2}\,,
\eeq
\[
\vA_{\bk} =\, \frac{i}{2
\pi^2}\,\frac{es}{2m}\;\frac{g(1-v^2)[\bk\times\si]+ \left(g -
2\gamma/(\gamma+1)\right)\,\bv(\bv\bk\si)}{k^2 - \ep(\om)\, \om^2}
\quad\quad\quad\quad\quad\quad\;
\]
\beq\label{At}
=\, \frac{i}{2 \pi^2}\,\frac{es}{2m}\;\frac{g(1-v^2)[(\bq +
\n\,\omega/v)\times\si]+\left(g -
2\gamma/(\gamma+1)\right)\,\bv(\bv\bq\si)}{q^2 - [\ep(\om)\,-
1/v^2]\,\om^2}\,.
\eeq

After substituting (\ref{pt}) and (\ref{At}) into (\ref{NH}) and
(\ref{EH}), we note that
\[
\int d^2 q \rightarrow \pi\int dq^2, \quad \int d^2 q \, q_m
\rightarrow 0\,, \quad \int d^2 q \, q_m \, q_n =
\frac{1}{2}\,\de_{mn}\, \pi\int dq^2\,q^2\,.
\]
We note also that
\[
\int^{\infty}_{-\infty}\frac{d\om \,\om\, q^2}{q^2 - [\ep(\om)\,-
1/v^2]\,\om^2} = \int^{\infty}_{-\infty} d\om \,\om \left\{1 +
\frac{[\ep(\om)\,- 1/v^2]\,\om^2}{q^2 - [\ep(\om)\,-
1/v^2]\,\om^2}\right\}
\]
\[
\quad \quad \quad \quad \quad \quad \quad =
\int^{\infty}_{-\infty}\frac{d\om \,\om^3 [\ep(\om)\,- 1/v^2]}{q^2
- [\ep(\om)\,- 1/v^2]\,\om^2}\,.
\]

Then the integral over $q^2$ is conveniently combined with all
explicit dependence on $\om$ into the following overall factor for
the spectral intensity:
\beq\label{dq}
I(\om)\, \sim \,f(\om)\,= - \,i\,\sum \,\om^3 \int\,\frac{d
q^2}{q^2 - [\ep(\om)\,- 1/v^2]\,\om^2}\,.
\eeq
The symbol $\,\sum\,$ in this expression means that one should sum
over the signs of the frequency: both $\om = +|\,\om|$ and $\om =
-|\,\om|$ contribute to the intensity $I(\om)$. All other
dependence of the total result on $\om$ is via $\ep(\om)$ only; in
our problem of the Cerenkov radiation, we restrict to the
frequencies corresponding to the region of transparency, i.e. to
real $\ep(\om)$ which is an even function of $\om$.

Let us analyze now expression
\[
f(\om) = -\,i\,\sum \,\om^3 \int\,\frac{d q^2}{q^2 - [\ep(\om)\,-
1/v^2]\,\om^2}\,
\]
entering result (\ref{dq}). The poles of its integrand correspond
obviously to the vanishing 4-momentum squared of a photon in the
medium.  Here, however, one should retain in $\ep(\om)$ its small
imaginary part: Im$\,\ep(\om) > 0$ for $\om > 0$, and
Im$\,\ep(\om) < 0$ for $\om < 0$. In other words, the poles of the integrand in
$f(\om)$ tend to the real axis from above for $\om > 0$, and from
below for $\om < 0$. Therefore, their contributions to the
integral are $i\pi$ and $-i\pi$, respectively. As to the real part
of the integral, it is an even function of $\om$ (together with
Re$\,\ep(\om)$), and therefore its contributions to the sum
$f(\om)$ cancel. Coming back to the poles, their contributions to
$f(\om)$ are $i\pi \om^3$ and $-i\pi (-\om)^3 = i\pi \om^3$, where from now on
$\om$ is positive. Thus,
\[
f(\om) = 2\pi \, \om^3,
\]

Then quite straightforward (though rather tedious) transformations
result in the following expressions for $(\bv\Na)\,\bH$ and
$(\bv\Na)\,\E$:
\[
(\bv\Na)\,\bH (\br = \bv t)\,=\, \frac{es}{2m}\,\frac{\om^3
\,d\om}{2v} \left\{-\si_\perp \left[\left(g - 2
+\frac{2}{\gamma}\right) \left(\ep - \frac{1}{v^2}\right) +\frac{2g}{\gamma^2
v^2}\right]\right.
\]
\beq
\left. - \si_\parallel \frac{g}{\gamma^2}\left(\ep -
\frac{1}{v^2}\right)\right\}\,,
\eeq
\beq
(\bv\Na)\,\E (\br = \bv t)\,=\, \frac{es}{2m}\,\frac{\om^3
\,d\om}{2v} \left[\left(g -
\frac{2\gamma}{\gamma+1}\right)\frac{1}{\ep}\left(\ep -
\frac{1}{v^2}\right) + 2g(1-v^2)\frac{1}{v^2}\right][\bv
\times\si_\perp]\,;
\eeq
here and below, $\si_\perp$ and $\si_\parallel$ are the components
of vector $\si$, orthogonal and parallel, respectively, to the
velocity $\bv$.

Now, plugging these expressions into (\ref{uto}) and (\ref{f}), we
arrive at the final general result for the spectral intensity of
Cherenkov radiation by a spinning particle:
\[
I(\om) d\om = \, \left(\frac{es}{2m}\right)^2\,\frac{\om^3
\,d\om}{2v}\left\{\left[\left(g - 2 +\frac{2}{\gamma}\right)^2
\left(n^2(\om) - \frac{1}{v^2}\right) - \left(g - 2 +
\frac{2}{\gamma+1}\right)^2\left(v^2
-\frac{1}{n^2(\om)}\right)\right.\right.
\]
\beq\label{gr}
\left.\left. +\frac{2g^2}{\gamma^4 v^2}\right]\si_\perp^2 +
\frac{g^2}{\gamma^3}\left(n^2(\om) -
\frac{1}{v^2}\right)\si_\parallel^2\right\}
\eeq

Few remarks on this result.

One should not bother about its formal singularity in $v$: anyway,
Cerenkov radiation takes place for $v \geq 1/n$ only.

Then, as distinct from the common Cerenkov radiation, here the
contribution to the energy loss due to $\si_\perp$ does not vanish
at the threshold, at $v = 1/n$.

At last, it is not exactly clear at first glance whether the
structure
\beq\label{qf}
\left(g - 2 +\frac{2}{\gamma}\right)^2 \left(n^2 -
\frac{1}{v^2}\right) - \left(g - 2 +
\frac{2}{\gamma+1}\right)^2\left(v^2 -\frac{1}{n^2}\right)
+\frac{2g^2}{\gamma^4 v^2}
\eeq
at $\si_\perp^2$ is positively definite (as it should be for
arbitrary $g$ and $\ga\,$!). To prove that this is the case indeed, we
note that the discussed quadratic function of $g$ is certainly
positively definite at $g \to \infty$ for $v \geq 1/n$. On the
other hand, the discriminant $d$ of this quadratic form is
negatively definite:
\[
d = - 4\ep\frac{v^2}{\ga^2}\left(1 -
\frac{1}{n^2v^2}\right)^2.
\]
So, quadratic form (\ref{qf}) is positively definite indeed.

Of course, in the case of a charged spinning particle the common
Cerenkov radiation takes place as well (and is strongly dominating
quantitatively). But don't we have then some combined effect, a Cerenkov-type
radiation of first order in spin? It
is practically obvious, by symmetry reasons, that such an
effect should not exist. But let us present somewhat more
quantitative arguments.
The effect could arise due to the Lorentz force $\F = e (\E + \bv
\times \bH )$, with $\E$ and $\bH$ generated by spin current
density (\ref{sto}). However, the magnetic contribution $e \bv
\times \bH$ to the energy loss $-\bv\F$ vanishes trivially. As to
the  corresponding electric contribution $-e\bv\E(\br = \bv t)$ to
the energy loss, one can demonstrate explicitly with formulae
(\ref{pt}), (\ref{At}) that it vanishes as well.
As explicitly one can demonstrate that the contribution to the
energy loss due to the Stern-Gerlach force (\ref{I0}), but now
with $\bH$ and $\E$ generated by the common convection current
$j_\mu(\br, t) = e (1,\, \bv)\,\de(\br-\bv t)$, vanishes as well.

\vspace{5mm}

{\bf 4.} In conclusion, let us consider some particular cases of
general result (\ref{gr}).

Let us start with a neutral particle with a finite magnetic moment
$\mu$. For $e \to 0$, $g \to \infty$,  and $\mu = esg/(2m)\to$
const, we obtain
\beq\label{I1}
I(\om)\,d\om\, =\,\frac{\mu^2\,\om^3}{2
v}\,d\om\,\left[\left(n^2 - \frac{1}{v^2} -v^2
+\frac{1}{n^2} + \frac{2}{\ga^4 v^2}\right)\si^2_\perp
+ \frac{1}{\ga^3}\left(n^2 -
\frac{1}{v^2}\right)\si_\parallel^2\right]\,.
\eeq
For $s = 1/2$ (e.g. for the Dirac neutrino with a mass and
magnetic moment), $\si^2_\perp = \si^2 - \sigma_z^2 = 2$ and
$(\si\n)^2 = \sigma_z^2 = 1$. So, here we obtain from (\ref{I1})
\beq\label{I2}
I(\om)\,d\om\,
=\frac{\mu^2\,\om^3}{v}\,d\om\,\,\left[\left(n^2 -
\frac{1}{v^2} -v^2 +\frac{1}{n^2}\right) +
\frac{1}{2\ga^3}\left(n^2 - \frac{1}{v^2}\right)+
\frac{2}{\ga^4 v^2}\right].
\eeq
In the classical limit, $s \gg 1$, radiation intensity (\ref{I1})
goes over into
\beq\label{I3}
I(\om)\,d\om\, =\,\frac{\mu^2\,\om^3}{2 v}\,d\om
\left[\left(n^2 - \frac{1}{v^2} -v^2 +\frac{1}{n^2} +
\frac{2}{\ga^4 v^2}\right)\sin^2\theta + \frac{1}{\ga^3}\left(n^2 -
\frac{1}{v^2}\right)\cos^2\theta\right],
\eeq
where $\theta$ is the angle between the spin and velocity.

The opposite limiting case is that of a charged particle with the
vanishing $g$-factor. The effect here is finite and looks as follows:
\beq\label{}
I(\om) d\om = \, \left(\frac{es}{2m}\right)^2\,\frac{2\om^3
\,d\om}{v}\left[\left(\frac{\gamma-1}{\gamma}\right)^2
\left(n^2 - \frac{1}{v^2}\right) - \left(
\frac{\gamma}{\gamma+1}\right)^2\left(v^2
-\frac{1}{n^2}\right) \right]\si^2_\perp.
\eeq

And at last let us mention the case $g=2$ (for instance, that of electron
if one neglects its small anomalous magnetic moment). Here
\[
I(\om) d\om = \, \left(\frac{es}{2m}\right)^2\,\frac{2\om^3
\,d\om}{v}\left\{\left[\frac{1}{\gamma^2}
\left(n^2 - \frac{1}{v^2}\right) -
\frac{1}{(\gamma+1)^2}\left(v^2
-\frac{1}{n^2}\right)+\frac{2}{\gamma^4 v^2}\right]\si_\perp^2\right.
\]
\beq
\left. + \frac{1}{\gamma^3}\left(n^2 -
\frac{1}{v^2}\right)\si_\parallel^2 \right\}.
\eeq

\begin{center} *** \end{center}
I am grateful to A.A. Pomeransky  for the interest to the work and
extremely useful discussions.

\noindent The work was supported in part by the Russian Foundation
for Basic Research through Grant No. 08-02-00960-a.

\end{document}